\documentclass[prl, aps, twocolumn, groupedaddress, superscriptaddress]{revtex4}
\usepackage{epsfig}
\usepackage{slashed}
\usepackage{graphicx}
\usepackage[usenames, dvipsnames]{color} 
\usepackage{psfrag}
\usepackage{graphics}
\usepackage{hyperref}
\usepackage{bm}
\usepackage{subfigure}

\begin{document}

\title{Rossby waves in rapidly rotating Bose-Einstein condensates}

\author{H. Ter\c{c}as}
\email{htercas@cfif.ist.utl.pt}
\affiliation{CFIF, Instituto Superior T\'{e}cnico, Av. Rovisco Pais 1, 1049-001 Lisboa, Portugal}
\author{J. P. A. Martins}
\affiliation{CGUL/IDL, University of Lisbon, Edif\'{i}cio C8, 1749-016 Lisboa, Portugal}
\author{J. T. Mendon\c{c}a}
\affiliation{CFIF, Instituto Superior T\'{e}cnico, Av. Rovisco Pais 1, 1049-001 Lisboa, Portugal}
\affiliation{IPFN, Instituto Superior T\'{e}cnico, Av. Rovisco Pais 1, 1049-001 Lisboa, Portugal}

\begin{abstract}
We predict and describe a new collective mode in rotating Bose-Einstein condensates, which is very similar to the Rossby waves in geophysics. In the regime of fast rotation, the Coriolis force dominates the dynamics and acts as a restoring force for acoustic-drift waves along the condensate. We derive a nonlinear equation that includes the effects of both the zero-point pressure and the anharmonicity of the trap. It is shown that such waves have negative phase speed, propagating in the opposite sense of the rotation. We discuss different equilibrium configurations and compare with those resulting from the Thomas-Fermi approximation.
\end{abstract}

\maketitle

The rotation of Bose-Einstein condensates (BEC) has attracted much attention recently, both theoretically and experimentally \cite{bloch, fetter}. Due to the superfluid character of the BEC, the effects of the rotation are quite different from those observed in a normal fluid and its properties strongly depend on the effects of the confinement. The pioneer experiments based in both the phase imprinting \cite{JILA} and the rotating leaser beam \cite{ENS, MIT} techniques independently confirmed the nucleation of quantized vortices, which is a clear manifestation of the superfluid properties of the condensate. Since then, much effort has been made to understand the dynamics of the rotating BEC \cite{pelster} and, in particular, the mechanisms of vortex nucleation \cite{jamaludin, sinha}. Particular interesting features of quantized vortices in Bose-Einstein condensates of alkali atoms are related to the formation of vortex arrays, where singly quantized vortices typically arrange in highly regular triangular lattices, similar to the Abrikosov lattice for supercondutors. Such a configuration is only possible when a sufficient amount of angular momentum is effectively transferred to the system, corresponding to a situation of a rapid rotation. The acquired angular velocity then tends to enlarge the rotating cloud and the centrifugal force is responsible for the flattening of the density profile towards a two-dimensional configuration. In the limit where the rotation frequency $\Omega$ approaches the transverse trapping frequency $\omega_\perp$, the quadratic centrifugal and the harmonic trapping potentials cancel out and the system is no longer bounded. The possibility of reaching high angular velocities is therefore provided by the addition of anharmonic terms to the trapping potential, making worthy the investigation of new equilibrium configurations with different vortex states and new collective modes \cite{cozzini1}. In the present work, we take advantage of such an interesting medium to predict a different hydrodynamical mode in rotating BECs, in complete analogy to the Rossby waves observed in geophysics. Rossby waves, also known as planetary waves, have been recognized for a long time as the main pattern of long period variability in the upper tropospheric winds \cite{rossby}. These are responsible for the well known cyclonic and anticyclonic systems that characterize the day-to-day weather systems in mid latitudes, and can be observed both in the upper troposphere and in the oceans. The waves exist due to the variation of the Coriolis parameter $f$ with latitude, which acts like a restoring force for an air particle that is disturbed from its equilibrium latitude. In a rotating BEC, the Coriolis parameter is replaced by twice the angular rotation frequency, $2\Omega$. We show that such waves are dispersive, their phase speed being always negative, which means that these oscillations always propagate westward with respect to the BEC rotation. We here derive a new equation for the Rossby waves in a rotating condensate, which accounts for the vortex lattice and the anharmonicity of the trap.\par

In the presence of a large number of vortices, Cozzini and Stringari showed that it is possible to average the velocity field over regions containing many vortex lines and assume that the vorticity is spread continuosly in the superfluid. This approximation is known as the diffused vorticity approach \cite{cozzini2} and simply corresponds to assume a rigid-body rotation $\mathbf{v}=\bm\Omega\times \mathbf{r}$, where the angular velocity is $\bm \Omega=\Omega \hat\mathbf{z}$ with $\Omega=\pi\hbar n_v/m$, $n_v(\mathbf{r})$ is the average vortex density in the vicinity of $\mathbf{r}$ and $m$ stands for the atomic mass. Therefore, the usual irrotationality condition $\bm \nabla \times \mathbf{v}=0$ is no longer valid and should be replaced by $\bm\nabla \times \mathbf{v}(\mathbf{r})=2\bm \Omega$. In that case, the macroscopic dynamics of the rotating BEC is provided by the rotational hydrodynamical equations in the rotating frame

\begin{equation}
\frac{\partial n}{\partial t}+\bm\nabla\cdot [n(\mathbf{v}-\bm{\Omega}\times \mathbf{r})]=0
\label{eq:3}
\end{equation}

\begin{equation}
\left(\frac{\partial}{\partial t}+\mathbf{v}\cdot\bm\nabla\right)\mathbf{v}=-gn-\bm\nabla V+\frac{\hbar^2}{2m^2}\bm\nabla \left(\frac{\nabla^2\sqrt{n}}{\sqrt{n}}\right)-2\bm\Omega\times \mathbf{v},
\label{eq:4}
\end{equation}
where $\mathbf{v}\cdot\bm\nabla\mathbf{v}=\bm \nabla (v^2)/2-\mathbf{v}\times(\mathbf{\Omega}\times\mathbf{v})$. The usual hydrodynamical calculations are based on the Thomas-Fermi approximation, which consists in neglecting the quantum pressure proportional to $\hbar^2$.  In this work, however, we include this quantum term, since we may be interested in Bogoliubov-like waves. This allows one to cast the effects of the quantum depletion in the condensate, which may be relevant to describe the so-called quantum turbulence \cite{kobayashi}. Here, $V(\mathbf{r},\Omega)=V_{trap}(\mathbf{r})-m\Omega^2 r^2/2$, with $r=(x^2+y^2)^{1/2}$, stands for the effective trapping potential, which reads

\begin{equation}
V(\mathbf{r},\Omega)=\frac{\hbar\omega_{\perp}}{2}\left[\left(1-\frac{\Omega^2}{\omega_\perp^2}\right)\frac{r^2}{a_{ho}^2}+\beta\frac{r^4}{a_{ho}^4}\right],
\label{pot}
\end{equation}
with $a_{ho}=\sqrt{\hbar/m\omega_\perp}$ being the characteristic harmonic oscillator length and $\beta$ the dimensionless anharmonicity parameter. The term $2\bm\Omega\times \mathbf{v}$ in Eq. (\ref{eq:4}) represents the Coriolis force, which will act as the restoring force for the oscillations considered here. We consider perturbations around the equilibirum configuration, making $n=n_0+\delta n$ and $\mathbf {v}=\delta \mathbf{v}$. In that case, the system can be described by the following set of perturbed equations

\begin{equation}
\frac{\partial}{\partial t}\delta n+\bm\nabla\cdot (n_0\delta \mathbf{v})=0
\label{eq:3b}
\end{equation}

\begin{equation}
\left(\frac{\partial}{\partial t}+\delta \mathbf{v}\cdot\bm\nabla\right)\delta\mathbf{ v}=-g\delta n-2\bm\Omega\times \delta \mathbf{v}+\frac{\hbar^2}{4m^2}\bm\nabla \left(\frac{\nabla^2\delta n}{n_\infty}\right),
\label{eq:4b}
\end{equation}
where $n_{\infty}$ is the peak density. The rotational velocity field can be split into to parts, $\mathbf{\delta v}\approx \delta\mathbf{v}_{0}+ \delta \mathbf {v}_{p}$, where

\begin{equation}
\delta\mathbf{v}_{0}=\frac{1}{2\Omega}\hat{\mathbf{z}}\times\mathbf{S}
\label{eq:5b}
\end{equation}
is the zeroth-order drift velocity, resulting from taking $d/dt=\partial/\partial t+\delta \mathbf{v}\cdot \bm{\nabla}=0$ in Eq. (\ref{eq:4b}) and 

\begin{equation}
\mathbf{S}=-g\bm\nabla \delta n+\frac{\hbar^2}{4m^2}\bm\nabla\left(\frac{\nabla^2\delta n}{n_\infty}\right).
\label{eq:5c}
\end{equation}
The polarization velocity $\delta \mathbf{v}_{p}$ is the first-order correction to the drift velocity (\ref{eq:5c}) and satisfies the following equation

\begin{equation}
\left(\frac{\partial}{\partial t}+\delta \mathbf{v}_{0}\cdot\bm\nabla \right)\delta\mathbf{v}_{0}=-2\bm\Omega\times \delta\mathbf{v}_{p},
\label{eq:6}
\end{equation}
which yields

\begin{equation}
\delta\mathbf{v}_{p}=-\frac{1}{4\Omega^2}\frac{\partial \mathbf{S_{\perp}}}{\partial t}-\frac{1}{8\Omega^3}\left(\mathbf{\hat z}\times\mathbf{S}\right)\cdot\bm{\nabla_{\perp}}\mathbf{S},
\label{eq:7}
\end{equation}
where $ \mathbf{S_{\perp}}=(S_{x},S_{y})$ is the transverse component of $\mathbf{S}$. The continuity equation (\ref{eq:3b}) can be written in the following fashion

\begin{equation}
\frac{d}{dt}\ln n+\bm\nabla\cdot\delta \mathbf{v}_{p}=0,
\label{eq:8}
\end{equation}
where the material derivative can be approximated as $d/dt\approx\partial/\partial t+\delta\mathbf{v}_{0}\cdot\bm{\nabla}$. Using the fact that $\ln n\approx \ln{n_0}+\phi$, where $\phi=\delta n/n_\infty$, and putting Eqs. (\ref{eq:5b}), (\ref{eq:7}) and (\ref{eq:8}) together, one should obtain

\begin{equation}
\left(1-r_{0}^2\nabla^2_{\perp}+\frac{1}{2}r_{0}^2\xi^2\nabla_{\perp}^4\right)\frac{\partial \phi}{\partial t}+2\Omega\left\{\psi,\phi-\nabla^2\psi+\ln n_{0}\right\}=0.
\label{eq:9}
\end{equation}
This equation is formally similar and generalizes Charney's equation \cite{charney}, also refereed in the literature as Charney-Hasegawa-Mima (CHM) equation. Here, $r_{0}=c_{s}/2\Omega$ represents the generalized Rossby radius, $c_{s}=\sqrt{gn_\infty/m}$ is the sound speed, $\xi=\hbar/\sqrt{2mgn_\infty}$ is the healing length \cite{pethick} and $\psi=r_{0}^2\phi-r_{0}^2\xi^2\nabla^2\phi/2$. The operator $\{a,b\}=r^{-1}(\partial_{r}a\partial_{\theta}b-\partial_{r}b\partial_{\theta}a)$ is simply the Poisson bracket in polar coordinates and $\theta$ represents the angular coordinate. The latter equation describes hydrodynamical drift waves in a rapidly rotating Bose-Einstein condensate and includes new features relatively to the CHM equation, widely used in the study of the dynamics of waves and turbulence in plasmas and in the atmosphere. Namely, the terms proportional to $\xi^2$ cast the effects of the zero-point pressure, which are known to play no role in geophysics. According to typical experimental conditions, we estimate the sound speed to be $c_{s}\sim 1$ mm/s, $\Omega \sim \omega_{\perp} \approx 2\pi \times 65$ Hz \cite{bretin}, a transverse harmonic oscillator radius of $a_{ho} \sim 1.7$ $\mu$m and a Rossby radius around $r_{0}\sim 1.2$ $\mu$m. The Rossby number, $\mbox{Ro}$, defines the ratio of the inertial to Coriolis forces   

\begin{equation}
\mbox{Ro}=\frac{c_{s}}{Lf},
\end{equation}
where $L$ is a typical length of the system and $f$ is the Coriolis parameter \cite{rossby, charney}. It characterizes the importance of Coriolis accelerations arising from planetary rotation and typically ranges from $0.1-1$ in the case of large-scale low-pressure atmospheres to $10^3$ in the case of tornados. Making $L=a_{ho}$ in a BEC, and $f=2\Omega$, we have $\mbox{Ro}=r_{0}/a_{ho}~\sim 0.7$, which suggests that the BEC can be regarded as a low-pressure atmosphere.\par
\begin{figure}
\includegraphics[scale=0.75]{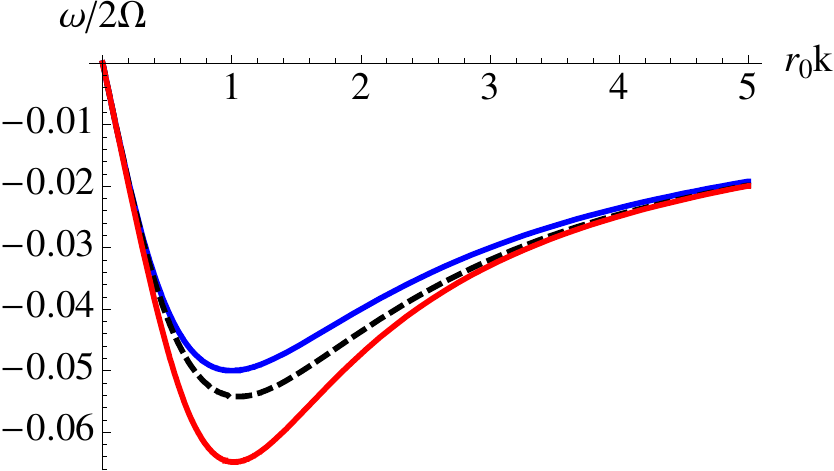}
\caption{(Color online) Dispersion relation of the Rossby waves in a rapidly rotating BEC, for $v_{R}=0.1c_{s}$. It is clearly shown that phase speed is generally negative. The blue full line corresponds to the Thomas-Fermi case,  $\xi=0$. The black dashed and red full lines respectively correspond to $\xi=0.7r_{0}$ and $\xi=1.3 r_{0}$.}
\label{fig1}
\end{figure}

 
We now show that a rotating BEC can sustain a new hydrodynamic mode corresponding to a drift-acoustic wave. For that purpose, we keep only the linear terms in Eq. (\ref{eq:9}), and look for perturbations of the form $\phi\sim e^{i(\mathbf{k}\cdot\mathbf{r}-\omega t)}$. The respective dispersion relation is then readily obtained and reads

\begin{equation}
\omega=-v_R k_{\theta}\frac{1+\xi^2k^2/2}{1+r_{0}^2k^2(1+\xi^2k^2/2)},
\label{disp}
\end{equation}
where $k_{\theta}=\mathbf{k}\cdot\mathbf{e}_{\theta}$ is the polar (or zonal) component of the wave vector $\mathbf{k}=(k_{x},k_{y})$. The term $v_R=-2\Omega r_0^2\partial_r \ln n_0$ is the generalized Rossby (drift) velocity. Because the equilibrium profile is generally very smooth, we expect $v_R$ to be small (compared to the Bogoliubov speed $c_{s}$), which suggest these waves to appear as a low frequency oscillation (compared to the both $\omega_\perp$ and $\Omega$). The dispersion relation (\ref{disp}) is similar to the expression for barotropic Rossby waves in the atmosphere \cite{rossby} and to the dispersion relation obtained for drift waves in a magnetized plasma \cite{mima}. For long wavelengths $r_{0}^2k^2\ll 1$ (and consequently $\xi^2k^2\ll 1$), Eq. (\ref{disp}) reduces to the {\it zonal flow} dispersion relation $\omega \approx -k_{\theta}v_{R}$. One of the remarkable features of the zonal, transverse acoustic, waves is that of having negative zonal phase and group velocities, $c_{\theta}^{(ph)}=c_{\theta}^{(g)}\approx -v_{R}$. It means that they propagate always ''weastward'' comparatively to the rotation of the condensate (which explains the negative values for the frequency in Eq.(\ref{disp})). For short wave lengths, one obtains the dispersion relation for the (actual) Rossby waves $\omega\approx -v_R k_{\theta}/\left(r_{0}^2k^2\right)$, with phase and group velocities approximately given by

\begin{eqnarray}
&&c_{\theta}^{(ph)}\approx -\frac{v_R}{r_{0}^2k^2} \nonumber\\
&&c_{\theta}^{(g)}\approx v_R \frac{2k_\theta/k-1}{r_0^2k^2},
\end{eqnarray}
In Fig. (\ref{fig1}), we plot the dispersion relation (\ref{disp}) for different values of the healing length $\xi$, using $v_R=0.1 c_s$. \par
Although a single Rossby wave of arbitrary amplitude is a solution of the linearized wave equation Eq. (\ref{disp}), a superposition of waves, generally, is not. The nonlinear interaction between the waves leads to a mechanism of energy transference. To study the interaction properties, one decomposes the solution into its Fourier series, $\phi_{\mathbf{k}}=\sum_{{\mathbf{k}}}\tilde\phi_{\mathbf{k}}\exp(i\mathbf{k}\cdot\mathbf{r})$ which, after plugging into Eq. (\ref{eq:9}), yields the following nonlinear equation

\begin{equation}
\frac{\partial\tilde \phi_{\mathbf{k}}}{\partial t}+i\omega_{k}\tilde \phi_{\mathbf{k}}=\sum_{\mathbf{k}_{1},\mathbf{k}_{2}}\Lambda_{\mathbf{k}_{1},\mathbf{k}_{2}}^{\mathbf{k}}\tilde\phi_{\mathbf{k}_{1}}\tilde\phi_{\mathbf{k}_{2}},
\label{nonlin}
\end{equation}
where 
\begin{eqnarray}
\Lambda_{\mathbf{k}_{1},\mathbf{k}_{2}}^{\mathbf{k}}&=&2r_{0}^2\delta\left(\mathbf{k}_{1}+\mathbf{k}_{2}-\mathbf{k}\right)\left(\mathbf{k}_{2}\times \mathbf{k}_{1}\right)\cdot\bm{\Omega}\nonumber\\
&\times& \frac{\left(1+\xi^2k_{1}^2/2\right)\left(1+r_{0}^2k_{2}^2+r_{0}^2\xi^2k_{2}^4/2\right)}{1+r_{0}^2k^2+r_{0}^2\xi^2k^4/2}
\end{eqnarray}
is the nonlinear coupling operator and $\omega_{k}$ is given by Eq. (\ref{disp}). Remarkably, only the waves that satisfy the condition $\mathbf{k}_{1}+\mathbf{k}_{2}=\mathbf{k}$ interact nonlinearly. The set of the waves satisfying this condition is known in the literature as the {\it resonant triad}. This resonance mechanism is able to transfer energy between different length scales, being one of the sources of classical turbulence in plasmas and in the atmosphere \cite{mima, pedlosky}. Here, due to the existence of additional terms that properly account for the quantum hydrodynamical features of the system, i.e, when large variations of the density profile are present, we believe that Eq. (\ref{nonlin}) may be used to describe turbulence in rotating Bose-Einstein condensates, opening a stage to explore the similarities and differences between classical and quantum turbulence.\par
\begin{figure}
\subfigure[]{
\includegraphics[scale=0.41]{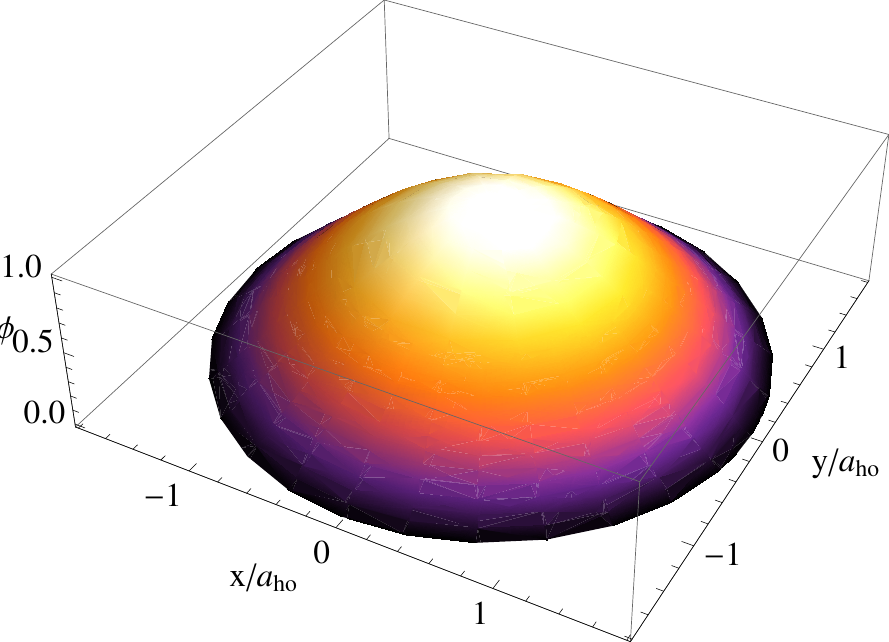}}
\subfigure[]{
\includegraphics[scale=0.41]{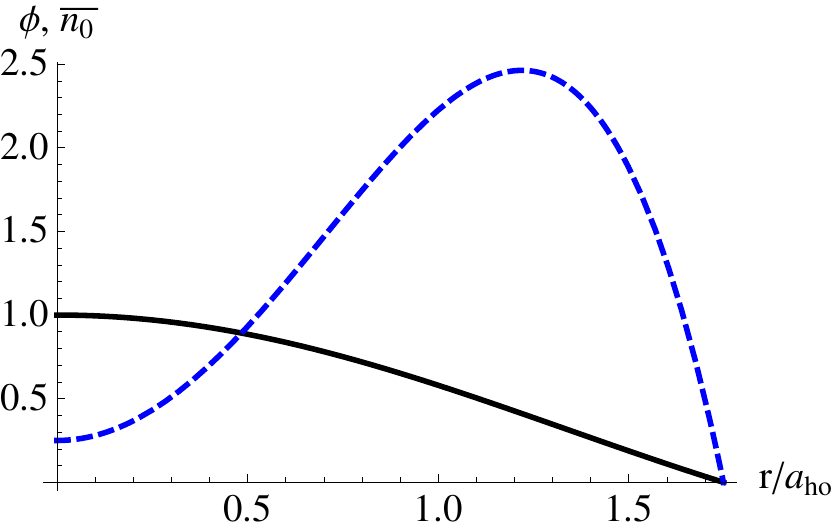}}
\subfigure[]{
\includegraphics[scale=0.41]{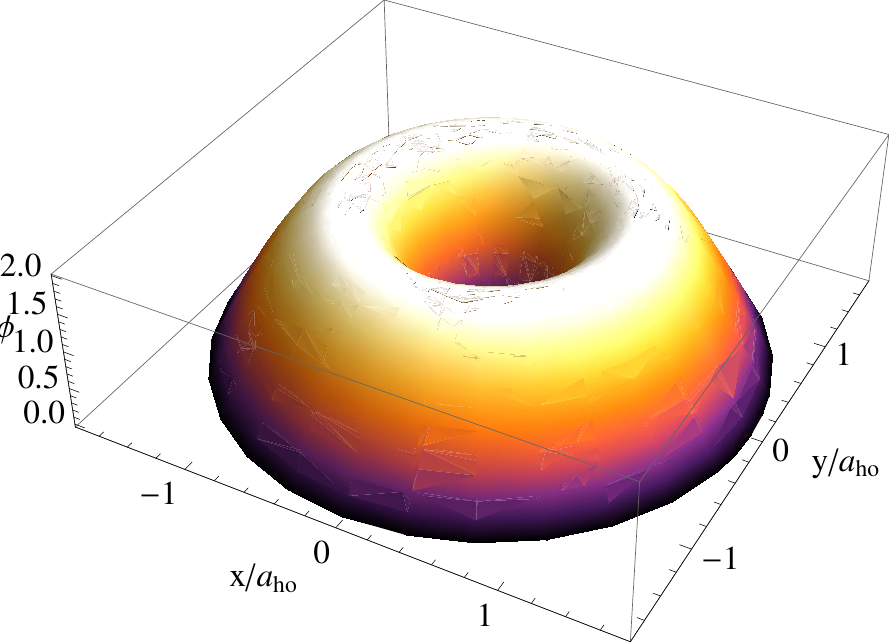}}
\subfigure[]{
\includegraphics[scale=0.41]{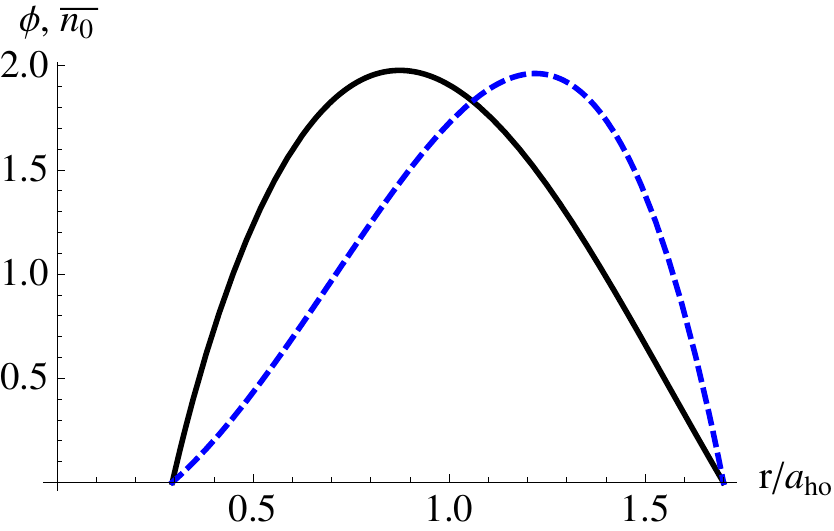}}
\caption{(Color online) Nonlinear stationary solitary wave resulting from the saturation of the triad resonant decay of Rossby waves, obtained for $\Omega=2.4\omega_\perp$ and $\beta=1.6$: a) $\mu=0.2\hbar\omega_\perp$  and c) $\mu=-0.2\hbar\omega_\perp$. Plots b) and d) respectively compare the radial structures (full lines) of a) and c) with the corresponding Thomas-Fermi equilibria (dashed lines) discussed in the text, obtained for the same set of parameters.}
\label{fig2}
\end{figure}

Another interesting feature of the Rossby waves in Bose-Einstein condensate is the possibility of finding localized structures, which may result, for example, from the saturation of the triad resonance mechanism mentioned above. Such purely nonlinear solitary structures can be obtained from the stationary solutions of Eq. {\ref{eq:9}}, which readily yields

\begin{equation}
\left(1+\frac{\xi^2}{2}\nabla_{\perp}^2 \right) \left\{\phi,\nabla_{\perp}^2\phi\right\}-\frac{\xi^2}{2}\left\{\phi,\nabla_{\perp}^4\phi\right\}=0.
\label{stationary}
\end{equation}
In the Thomas-Fermi limit, the latter expression simply reduces to $\{\phi,\nabla_{\perp}^2\phi\}=0$, which is satisfied for a family of functions $\nabla_{\perp}^2\phi=\mathcal{F}(\phi)$, where $\mathcal{F}(x)$ is an arbitrary function of its argument. The different choices for $\mathcal{F}$ will lead to different structures, which describe many physical nonlinear stationary solutions. For example, for the choice $\mathcal{F}(\phi)\propto\exp(-2\phi)$, Stuart \cite{stuart} showed that the so-called ''cat-eye'' solution describes a vortex chain in a magnetized plasma sheet, which has been observed experimentally in mixing layer experiments \cite{browand}. However, in the present case, there are physical limitations that impose specific constraints to the choice of the solutions. In particular, as discussed in \cite{cozzini1}, the equilibrium profile associated with the potential in Eq. (\ref{pot}), which is given by $n_{0}(r)=n_{\infty}(R_{+}^2-r^2)(r^2-R_{-}^2)$ (the peak density is $n_{\infty}=\beta\hbar\omega_{\perp}/2g$), must vanish at the Thomas-Fermi radii defined as follows

\begin{equation}
\frac{R_{\pm}^2}{a_{ho}^2}=\frac{\Omega^2-\omega_{\perp}^2}{2\beta\omega_{\perp}^2} \pm \sqrt{\left(\frac{\Omega^2-\omega_{\perp}^2}{2\beta\omega_{\perp}^2}\right)^2+\frac{2\mu}{\beta\hbar\omega_{\perp}}},
\end{equation} 
where $\mu$ represents the chemical potential. For $\mu>0$, the radius $R_{-}$ is purely imaginary and the density vanishes at $R=R_{+}$, while for $\mu<0$ both $R_{-}$ and $R_{+}$ are present. This reflects the transition occurring at $\mu=0$, where a hole forms in the centre of the condensate and the equilibrium profile assumes an annular shape. The simplest nonlinear structure that verifies such constraints is obtained for $\mathcal{F}(\phi)=-\kappa\phi$, and the respective radial solution, for $\ell=0$, yields $\phi(r)=AJ_0(\kappa r)+ BY_0(\kappa r)$. The values of $\kappa$, $A$ and $B$ are such that the solution vanishes at the radii $R_\pm$. In Fig. (\ref{fig2}), we plot two possible solitary structures in the overcritical rotation regime $\Omega>\omega_\perp$, obtained for both $\mu>0$ and $\mu<0$. It is interesting to observe that, even for the same set of parameters, the resulting solitary structures may differ from the usual Thomas-Fermi equilibrium profiles discussed above.\par
In the spirit of a mean field description, we derived an equation that governs the dynamics of a new drift mode in rotating Bose-Einstein condensates, which is in close analogy to the Rossby waves in geophysics. Our equation, which has been established for the first time in the context of superfluids (for the best of our knowledge), casts the effect of the anharmonicity of the trap, and can thus be extended to the overcritical rotating regime. After linearization, we derived the dispersion relation for the Rossby waves and showed that they propagate on the opposite sense of that defined by the angular rotating frequency $\bm{\Omega}$. A particular feature of these waves is that they resonantly decay in form of triads, which is a clear manifestation of a three-wave mixing mechanism in BEC. A very recent and interesting work by Bludov et al. \cite{bludov} also establishes the connection between Bose-Einstein condensates and geophysics, where the authors report on the occurrence of rogue waves, a well known phenomenon in deep oceans. We therefore believe that future numerical and experimental work would reveal interesting features concerned to the nonlinear dynamics in these systems, with special emphasis to the issue of turbulence spectrum, where we believe that the numerical integration of Eq. (\ref{nonlin}) would be relevant. A more detailed investigation of the temporal evolution of the Rossby wave turbulence, as well as the stability of the solitary structures certainly deserve further attention.\par
This work was partly supported by Funda\c{c}\~{a}o para a Ci\^encia e Tecnologia (FCT-Portugal) through the grants number SFRH/BD/37452/2007 and SFRH/BD/37800/2007.

\bigskip


\begin{thebibliography}{10}
\bibitem{bloch} I. Bloch, J. Dalibard and W. Zweger, Rev. Mod. Phys. {\bf 80}, 885 (2008). 
\bibitem{fetter} A. Fetter, Rev. Mod. Phys. {\bf 81}, 647 (2009). 
\bibitem{JILA} M. R. Matthews, B. P. Anderson, P. C. Haljan, D. S. Hall, C. E. Wieman, and E. A. Cornell, Phys. Rev. Lett. {\bf 83}, 2498 (1999).
\bibitem{ENS} K. W. Madison, F. Chevy, W. Wohlleben, and J. Dalibard, Phys. Rev. Lett. {\bf 84}, 806 (2000).
\bibitem{MIT} J. R. Abo-Shaeer, C. Raman, J. M. Vogels, and W. Ketterle, Science {\bf 292}, 476 (2001).
\bibitem{pelster} S. Kling and A. Pelster, Phys. Cold. Trap. Atoms {\bf 19}, 1072 (2009).
\bibitem{sinha} S. Sinha and Y. Castin, Phys. Rev. Lett {\bf 87}, 190402 (2001).
\bibitem{jamaludin} N. A. Jamaludin, N. G. Parker and A. M. Martin, Phys. Rev. A {\bf 77}, 051603(R) (2008).
\bibitem{cozzini1} M. Cozzini, A. L. Fetter, B. Jackson and S. Stringari, Phys. Rev. Lett. {\bf 94}, 100402 (2005).
\bibitem{rossby} C.-G. Rossby et al., Journal of Marine Research {\bf 1}, 38 (1939).
\bibitem{cozzini2} M. Cozzini and S. Stringari, Phys. Rev. A {\bf 67}, 041602(R) (2003).
\bibitem{kobayashi} M. Kobayashi and M. Tsubota, J. Low Temp. Phys. {\bf 150}, 587 (2007).
\bibitem{charney} J. G. Charney and A. Elliassen, Tellus {\bf 1}, 38 (1949).
\bibitem{pethick} C. J. Pethick and H. Smith, {\it Bose-Einstein Condensation in Dilute Gases}, 2nd Ed. (Cambridge Univ. Press, Cambridge, 2008). 
\bibitem{bretin} V. Bretin, S. Stock, Y. Seurin and J. Dalibard, Phys. Rev. Lett. {\bf 80}, 885 (1998).
\bibitem{mima} A. Hasegawa, K. Mima, Phys. Fluid. {\bf 21},  87 (1978).
\bibitem{pedlosky} J. Pedlosky, {it Geophysical Fluid Dynamics}, 2nd Ed., Springer-Verlag (1986).
\bibitem{stuart} J. T. Stuart, J. Fluid. Mech. {\bf 29}, 417 (1967).
\bibitem{browand} F. K. Browand and P. D. Wiedman,  J. Fluid Mech. {\bf 76}, 127, (1976).
\bibitem{bludov} Yu. V. Bludov, V. V. Konotop and N. Akhmediev, Phys. Rev. A {\bf 80}, 033610 (2009). 
\end{thebibliography}
\end{document}